\documentstyle[preprint,aps]{revtex}
\widetext
\draft
\tighten

\begin{document}

\title{On the theoretical possibility of the electromagnetic  scalar
potential wave spreading with an arbitrary velocity in vacuum}

\bigskip

\author{{\bf Andrew E. Chubykalo and   Vladimir V. Onoochin\thanks
{``Sirius", Nikoloyamski lane 3A, Moscow, 109004, Russia.  }}}

\address {Escuela de F\'{\i}sica, Universidad Aut\'onoma de Zacatecas \\
Apartado Postal C-580\, Zacatecas 98068, ZAC., M\'exico\\
e-mails: {\rm andrew@logicnet.com.mx},   and
{\rm andrew@ahobon.reduaz.mx}}

\date{\today}

\maketitle

\baselineskip 7mm
\bigskip
\bigskip
\bigskip
\bigskip

\pacs{PACS numbers: 03.50.-z, 03.50.De}

% \clearpage
\begin{abstract}
In this work we revisit the process of constructing  wave
equations for the scalar and vector potentials of an electromagnetic
field, and show that a wave equation with an arbitrary velocity (including
a velocity higher than the velocity of light in vacuum) for the scalar
potential exists in the framework of classical electrodynamics.  Some
consequences of this fact are considered.  \end{abstract}

Our letter is devoted to the discussion of a possibility of the existence
of sub- and superluminal electromagnetic waves in  vacuum.

A considerable number of experimental and theoretical works about
superluminal spreading of electromagnetic waves, particles and other
objects have recently been published, as  mentioned by E. Recami in
[1,2], Walker in [3], Kotel'nikov in [4,5]  and in the book edited by
Chubykalo {\it et al} [6] (see also references in the mentioned works). J.
Marangos wrote in his brilliant note ``Faster than a speeding photon" [7]:
``{\sl The textbooks say nothing can travel faster than light, not even
light itself.  New experiments show that is no longer true, raising
questions about the maximum speed at which we can send information.}" (see
also the bibliography in this work). Really, a series of recent
experiments, performed at Cologne[8], Berkeley[9], Florence[10] and
Vienna[11], and quite recent experiments by W. Tittel {\it et al} [12]
revealed that evanescent waves (in undersized waveguides, e.g.) seem to
spread with a superluminal group velocity. For example, in up-to-date
experiments by Mugnai {\it et al} (see their work [13])
superluminal behavior in the propagation of microwaves (centimeter
 wavelength) over much longer distances (tens of centimeters) at a speed
7\% faster than $c$  was reported.

In the majority of cases, these works almost directly
declare that  generally accepted electrodynamics must be sufficiently
reconsidered.\footnote{See incidentally ``Essay on non-Maxwellian
theories of electromagnetism" by V. Dvoeglazov [14].}

In this paper we would like to address the problem of electromagnetic
waves spreading with an {\it arbitrary}  velocity in vacuum ($v\leq c$ and
$v\geq c$) conditioned by {\it a choice of gauges}.  In  other words, we
attempt to explain here these superluminal electromagnetic phenomena using
well-known arbitrariness of a choice of gauges permitted by classical
 electrodynamics.

 Is there a way of showing a possibility of the existence of the
superluminal wave processes in vacuum mentioned in the beginning of the
letter, without leaving the framework of classical electrodynamics?  Here
we show that, yes, there is.

It is known, if we are given potentials ${\bf A}$ and $\varphi$, then
these uniquely determine the fields ${\bf E}$ and ${\bf H}$:

\begin{equation}
{\bf E}=-\nabla\varphi-\frac{1}{c}\frac{\partial{\bf A}}{\partial t}\qquad
{\rm and}\qquad {\bf H} = \nabla\times{\bf A}.
\end{equation}
However, it is well known that the same field can correspond to different
potentials. Electric and magnetic fields determined from equations (1)
actually do not change upon replacement of ${\bf A}$ and $\varphi$ by
${\bf A}^{\prime}$ and $\varphi^{\prime}$, defined by

\begin{equation}
{\bf A}^{\prime}={\bf A}+\nabla f\qquad {\rm and}\qquad
\varphi^{\prime}=\varphi-\frac{1}{c}\frac{\partial f}{\partial t}.
\end{equation}

Only those quantities  which are invariant with
respect to the transformation (2) have physical meaning, so all equations
must be invariant under this transformation.

This non-uniqueness of the potentials gives us
the possibility of choosing them so that they fulfill one auxiliary
condition (gauge) chosen by us. It means that we can set one, since we may
choose the function $f$ in (2) arbitrarily.

Let us recall how a necessity of choosing a gauge condition arises.
One substitutes fields ${\bf E}$ and ${\bf H}$, expressing by potentials
${\bf A}$ and $\varphi$, into Maxwell equations

\begin{eqnarray} && \nabla\cdot{\bf
E}=4\pi\varrho,\\ && \nabla\cdot{\bf H}=0,\\ && \nabla\times{\bf
H}=\frac{4\pi}{c}{\bf j}+\frac{1}{c}\frac{\partial {\bf E}}{\partial t},\\
&& \nabla\times{\bf E}=-\frac{1}{c}\frac{\partial {\bf H}}{\partial t},
\end{eqnarray}
to obtain equations which are  easier-to-use than the original equations
with respect to fields ${\bf E}$ and ${\bf H}$.  After direct substituting
(1) into Maxwell equations we have:

\begin{equation}
\Delta\varphi+\frac{1}{c}\frac{\partial}{\partial t}{\rm div}\,{\bf A}
=-4\pi\varrho
\end{equation}
and
\begin{equation}
\Delta{\bf A}-\frac{1}{c^2}\frac{{\partial}^2{\bf A}}{\partial t^2}
-{\rm grad}\left({\rm div}\,{\bf A}+\frac{1}{c}\frac{\partial\varphi}
{\partial t}\right)=-\frac{4\pi}{c}{\bf j}.
\end{equation}

Thus because of non-uniqueness of the potentials, we can always subject
them to an auxiliary condition. For this reason, we try to choose  this
condition so as the system of equations (7) and (8) (or at least one of
them) would be transformed into some ``easy-to-solve" equations. Let us
consider this problem from a purely formal, mathematical point of view.
Note if a certain connection between ${\rm div}\,{\bf A}$ and
$\frac{\partial\varphi} {\partial t}$ exists then because of
dimensions condition this connection must generally look like

\begin{equation}
{\rm div}\,{\bf A}
+\alpha\frac{\partial\varphi}{\partial t}=0,
\end{equation}
where $\alpha$ is an {\it arbitrary} constant parameter with  dimensions
$\left[\frac{1}{{\tt cm/sec}}\right]$. We will show that the condition (9)
can satisfy (2) by a corresponding choice of function $f$ in (2). For this
we substitute values ${\bf A}^{\prime}$ and $\varphi^{\prime}$ into (9):

\begin{equation}
\nabla{\bf A}+\Delta f+\alpha\frac{\partial\varphi}{\partial t}-
\frac{\alpha}{c}\frac{\partial^2 f}{\partial t^2}=0.
\end{equation}
Let us make now a formal conjecture that a perturbation of potential
$\varphi$ spreads with a certain {\it arbitrary}
constant velocity $v$ (for a given process)  which is not necessarily
equal to $c$.  We choose the arbitrary constant $\alpha$ as
$\alpha=\frac{c}{v^2}$ then the condition (9) becomes

\begin{equation}
{\rm div}\,{\bf A}
+\frac{c}{v^2}\frac{\partial\varphi}{\partial t}=0,
\end{equation}.

 Consequently from (10) we obtain the equation for $f$:
 \begin{equation}
 \Delta f-\frac{1}{v^2}\frac{\partial^2 f}{\partial t^2}=
 F({\bf r},t),
 \end{equation}
 where $F({\bf r},t)=-\nabla{\bf A}-\frac{c}{v^2}
 \frac{\partial\varphi}{\partial t}$ is a given function ${\bf r}$ and
 $t$. Substituting the function $f$ from a solution of Eq. (12) into
 formulas (2) we find values of potentials ${\bf A}^{\prime}$ and
 $\varphi^{\prime}$ satisfying {\it gauge} (11)\footnote{note that if we
 choose the arbitrary constant $\alpha$ as $\alpha=\frac{1}{c}$ instead of
  gauge (11), we obtain the well-known Lorentz gauge $$ {\rm div}\,{\bf A}
+\frac{1}{c}\frac{\partial\varphi}{\partial t}=0.$$}.

The gauge (11) unlike the Lorentz gauge, has no relativistically invariant
character. But  gauge (11) is obviously allowable in classical
electrodynamics {\it on parity with other} well-known gauges: the Coulomb
gauge

\begin{equation}
 {\rm div}\,{\bf A}=0\qquad{\rm and}\qquad \Delta\varphi=- 4\pi\varrho
\end{equation}
and the so-called radiation gauge
\begin{equation}
 {\rm div}\,{\bf A}=0\qquad{\rm and}\qquad \varphi= 0,
\end{equation}
which in turn also have no relativistically invariant character [15].

Let us apply the gauge (11) to Eqs. (7) and (8). As a result we obtain two
equations:

\begin{equation}
\Delta\varphi-\frac{1}{v^2}\frac{\partial^2\varphi}{\partial t^2} =
-4\pi\varrho
\end{equation}
and
\begin{equation}
\Delta{\bf A}-\frac{1}{c^2}\frac{\partial^2{\bf A}}{\partial t^2}=
{\rm grad}\left(\frac{v^2-c^2}{cv^2}\frac{\partial\varphi}{\partial
t}\right)-\frac{4\pi}{c}{\bf j}.
\end{equation}

In current- and charge-free regions these equations become
\begin{equation}
\Delta\varphi-\frac{1}{v^2}\frac{\partial^2\varphi}{\partial t^2} =
0
\end{equation}
and
\begin{equation}
\Delta{\bf A}-\frac{1}{c^2}\frac{\partial^2{\bf A}}{\partial t^2}=
\left(\frac{v^2-c^2}{cv^2}\right){\rm grad}\frac{\partial\varphi}{\partial
t}.
\end{equation}

We can see that the equations obtained differ strongly  from the
well-known wave equations for charge-free space in the Lorentz gauge

\begin{equation}
\Delta\varphi_L-\frac{1}{c^2}\frac{\partial^2\varphi_L}{\partial t^2} =
0
\end{equation}
and
\begin{equation}
\Delta{\bf A}_L-\frac{1}{c^2}\frac{\partial^2{\bf A}_L}{\partial t^2}=
0
\end{equation}
(index $L$ denotes that these potentials obey Lorentz gauge). Actually, a
solution of Eq.(17)  is a wave of perturbation  of the scalar potential
spreading in vacuum with the arbitrary phase velocity\footnote{recall
that in vacuum, the phase velocity and the group velocity of
electromagnetic perturbation spreading coincide.} $v$ which can be both
lower than $c$ and higher than $c$ (unlike the solution of Eq.(19)!) Then
there is Eq.(18) which is a wave equation with a source (unlike Eq.(20))
despite the fact that this equation is written for charge(current-)-free
space.  The function of source in Eq.(18) is a gradient of changing of the
solution of Eq.(17) with changing of time. Note that the gauge (11) is
more general gauge than Coulomb and Lorentz gauges.  Actually, if we
choose the arbitrary constant velocity $v$ in (11) as $c$ we obtain the
Lorentz gauge and corresponding wave equations for potentials (19),(20),
on the other hand if $v$ in (11) tends to infinity we immediately obtain
the Coulomb gauge and corresponding equations for potentials:

\begin{equation}
\Delta\varphi_C =
-4\pi\varrho
\end{equation}
and
\begin{equation}
\Delta{\bf A}_C-\frac{1}{c^2}\frac{\partial^2{\bf A}_C}{\partial t^2}=
{\rm grad}\left(\frac{1}{c}\frac{\partial\varphi_C}{\partial
t}\right)  -\frac{4\pi}{c}{\bf j}.
\end{equation}
(index $C$ denotes that these potentials obey Coulomb gauge).

Obviously, equations (15)-(18) and specially Eqs. (15) and (17) can lead
the way to found a theory of superluminal electromagnetic interactions and
signal transfer in vacuum, which experimental evidences were mentioned in
the Introduction. And the scalar potential $\varphi$ from Eqs. (17)
has to play the leading role in constructing this theory.

Against this, one can say  that there is the well-known, generally accepted
opinion in classical electrodynamics that electromagnetic potentials are
just auxiliary quantities which have no  real physical meaning (unlike
real mensurable magnitudes ${\bf E}$ and ${\bf H}$). We can reply to this
retort as follows:

On the one hand,  from {\it generally accepted} classical
electrodynamics we know that the Poynting vector is {\it proportional} to
the density of the electromagnetic field momentum. But on the other hand,
paradoxes connected with the Poynting vector (and, correspondingly, with
an  energy and momentum distribution) exist and they are well-known.  For
example, in [16] it is noted:  if a point charge $Q$ is vibrating in some
mechanical way along the $X$-axis with respect to a certain point $x_Q$,
then the value of electromagnetic energy density $w$ (which is a point
function like ${\bf E}$) on the same axis will be also oscillating.
Immediately the question arises:  {\it how does the test charge $q$ at the
point of observation, lying at some fixed distance from the point $x_Q$
along the continuation of the X-axis, ``know" about the charge $Q$
vibration}?  In other words, we have a rather strange situation:  the
Poynting vector ${\bf S}=\frac{c}{4\pi}[{\bf E}\times{\bf H}]$ is zero
along this axis\footnote{Note that recently many works devoted to
the Poynting vector concept was published. See, e.g., [17] and
corresponding references there.} (because {\bf H} is zero along this line)
but the energy and the momentum, obviously ``pass" from point to point
along this axis.  This means that {\it we cannot be sure any more} that
exclusively using the vector fields ${\bf E}$ and ${\bf H}$ allows us to
characterize the process of distributing the electromagnetic energy and
momentum in vacuum within the framework of classical electrodynamics.  It
would be very interesting to carry out thorough research in this problem.
At this stage, we can only say that it is unlikely that one can solve this
problem without taking into account a {\it physical reality} of
electromagnetic potentials waves. On the other hand the equation (17) and
(18) are conjunct, i.e. the radiator (source) in the wave equation (18)
for the vector potential is the function of the superluminal (for example)
solution of (17)! What does it mean theoretically? It means that {\it the
source of the vector potential's waves spreads with an superluminal
velocity}. But the vector potential in turn produces the variable magnetic
field (it follows from $\nabla\times{\bf A}={\bf H}$). The variable
magnetic field in turn produces the variable electric field. It means that
a perturbation of the scalar potential started up in some distant point
reaches the given point and produces in it {\it mensurable} fields ${\bf
E}$ and ${\bf H}$.

It is an accepted truth that, for the purpose of long-range radio
communications, one employs mainly transverse electromagnetic waves
propagating at the speed of light. These waves are described
mathematically by relatively simple equations whose consequences can be
easily verified by experiment. On the other hand, as it was noted in
the recent work [18], there exists an extensive area of electromagnetic
phenomena characterized by a complicated and mainly approximate
mathematical formulation whose consequences cannot be easily tested
experimentally\footnote{Optical phenomena are not excluded, since we
know nothing about the details of quantum transitions responsible for the
emission of photons by atoms.}.  This is precisely the area where one can
expect to find unknown effects and unusual properties of the
electromagnetic field. Such an expectation is supported by the fact that
even  classical electrodynamics does not actually prohibit either the
superluminal velocity or the longitudinal electromagnetic waves if we
take into account {\it the physical reality} of potentials and existence of
Eqs. (15)-(18).  It is clear, therefore, that in general electromagnetic
fields may propagate in a manner drastically different from that with
which we are familiar on the basis of radio broadcasting.

For objectiveness we have to mention that many works are published where
authors claim that the absolute majority of experiments discovered the
superluminal propagation are incorrect (see, for example, the book [6] the
Section ``Contra"). But we just would like to note that it cannot be
emphasized enough that in the problem of the experimental confirmation or
confutation of the superluminal propagation the last word has not yet been
said on this matter. But now we have the {\it formal proof}, based only on
Maxwell electrodynamics, that electromagnetic waves connected with a
perturbation of scalar potential $\varphi$ (Eqs.  (15), (17)) {\it can}
travel either faster than the speed of light in vacuum $c$ or slower than
$c$ but not exclusively with $c$ exactly! Here let us observe (following
E.Recami [19]) that the particular role of the speed of light in the
Special Relativity is due to its {\it invariance}, and not to the fact
that it is (or it is not) the {\it maximal} one.

\acknowledgments

The authors would like to express their gratitude to Profs. Valeri
Dvoeglazov and Stanislav Pavlov from Zacatecas University for their
discussions and critical comments.  We would also like to thank Annamaria
D'Amore for revising the manuscript.\\

\begin{center}
{\large Appendix} \footnote{This Chapter is added after sending the
manuscript}
\end{center}

Recently, two works [20, 21] on the subject considered in this paper appeared,
where it is stated that despite of non-retarded origin of the scalar potential
in any gauge ({\it except the Lorentz one}), the {\bf E} field calculated in
{\it any gauge} is retarded. To the authors' point of view, it is not so.

One can easily see that in both cited works \footnote{We focus on the
Coulomb gauge as it is done in [20, 21], but basically it can be shown for any
gauge}, the key point of proof of equivalence of the gauges is to show that
the current density ${\bf j}$ can be presented as a sum of two currents,
longitudinal and transversal:

\begin{equation}
{\bf j}_{\vert \vert } + {\bf j}_\bot = {\bf j}
\end{equation}
by the way, the longitudinal current, which is
the source in the equation for the vector potential, is yielded by the
{\it action-at-distant} scalar potential so it is described by {\it highly
nonlocal function} (Eq. (3.9) of [21] and Eq. (2.8) of [20]).
\begin{equation}
{\bf j}_{\vert \vert } = - \frac{1}{4\pi }\nabla \int{
\frac{\nabla ' \cdot {\bf j}({\bf r}',t)}{\left| {\bf r} - {\bf r}'\right|}}
d{\bf r}'
\end{equation}
Correspondingly, the transversal current is described by highly nonlocal
function too:

\begin{equation}
{\bf j}_\bot = \frac{1}{4\pi }\nabla \times \nabla \times \int
{\frac{{\bf j}({\bf r}',t)}{\left| {\bf r} - {\bf r}' \right|}} dr'
\end{equation}
Eqs. (24) and (25) correspond to Eqs. (6.49) and (6.50) of [22], where the
proof that the sum of \textit{rhs} of these equations is equal to the
\textit{rhs} of Eq. (23) is given. This proof is based on application of the
vector identity

\begin{equation}
\label{eq47}
\nabla \times \nabla \times {\bf j} = \nabla ( \nabla \cdot {\bf j} ) -
\nabla ^2{\bf j}
\end{equation}

\noindent
and equation $\nabla ^2(1 / \left| {\bf r} - {\bf r}' \right|) = - 4\pi
\delta ({\bf r} - {\bf r}')$ . However, it should be noted that while
proving validity of Eq. (23), (or Eq. (6.48) of [22]), two formal rules are
broken here, {\it i.e.}

\begin{itemize}
\item{In Eq. (25) the differential operator $\nabla \times \nabla \times$
must act on the vector too but $1 / \left| {\bf r} - {\bf r}' \right|$
is a scalar,}

\item{the differential operator $\nabla \times \nabla \times$ acts on
external variable but the vector equality (26) must be applied to internal
variables}.
\end{itemize}

So, the proof given in [22] is incorrect and, therefore, calculations of
works [20, 21] based on this proof are incorrect too.

Physical explanation of incorrectness of this proof is simple: original
current density is the {\it local} quantity. After Eq. (23), it is proposed
to form the local quantity from two nonlocal quantities. Let us consider a
point of space where the current is zero. We should create from this {\it
zero} two vector fields; by the way one of them is {\it rotational} (Eq. 25))
and second is {\it irrotational} (Eq. (24)), {\it i.e.} two fields different
in their origin. So the rotational field is equal to 'minus' irrotational
field and, because for the vector the sign "minus" correspond to opposite
direction of the vector, {\it rotational field = - irrotational field}, that
is nonsense.

The reader can see that we do not use Eq. (23) in our calculations and
just in this point, there is {\it essential difference} between this work
and all previous works on the gauges in the classical electrodynamics where 
correctness of Eq. (23) is undeniably accepted.

\end{document}